\documentclass
[prc,twocolumn,showpacs,preprintnumbers,amsmath,amssymb,superscriptaddress,floatfix,nofootinbib]{revtex4}
\usepackage{graphicx}
\usepackage{amsmath}
\usepackage{amsfonts}
\usepackage{amssymb}%
\begin{document}
\title{Plausible explanation of the $\Delta_{5/2^{+}}(2000)$ puzzle}
\author{Ju-Jun Xie}
\email{xiejujun@ific.uv.es}
\affiliation{Instituto de F\'\i sica Corpuscular (IFIC), Centro Mixto CSIC-Universidad de
Valencia, Institutos de Investigaci\'on de Paterna, Aptd. 22085, E-46071
Valencia, Spain}
\affiliation{Department of Physics, Zhengzhou University, Zhengzhou, Henan 450001, China}
\author{A. Mart\'inez Torres}
\email{amartine@yukawa.kyoto-u.ac.jp}
\affiliation{Yukawa Institute for Theoretical Physics, Kyoto University, Kyoto 606-8502, Japan}
\author{E. Oset}
\email{oset@ific.uv.es}
\author{P. Gonz\'alez}
\email{pedro.gonzalez@uv.es}
 \affiliation{Instituto de F\'\i sica Corpuscular (IFIC), Centro Mixto CSIC-Universidad de Valencia,
Institutos de Investigaci\'on de Paterna, Aptd. 22085, E-46071
Valencia, Spain} \affiliation{Departamento de F\'{\i}sica
Te\'{o}rica, Universidad de Valencia, Valencia, Spain.}

\preprint{YITP-10-109}

\begin{abstract}
From a Faddeev calculation for the
$\pi-(\Delta\rho)_{N_{5/2^{-}}(1675)}$ system we show the plausible
existence of three dynamically generated $I(J^{P})=3/2~(5/2^{+})$
baryon states below 2.3 GeV whereas only two resonances,
$\Delta_{5/2^{+}}(1905)(\ast\ast\ast\ast)$ and
$\Delta_{5/2^{+}}(2000)(\ast\ast),$ are cataloged in the Particle
Data Book Review. Our results give theoretical support to data
analyses extracting two distinctive resonances,
$\Delta_{5/2^{+}}(\sim1740)$ and $\Delta_{5/2^{+}}(\sim2200),$ from
which the mass of $\Delta_{5/2^{+}}(2000)(\ast\ast)$ is estimated.
We propose that these two resonances should be cataloged instead of
$\Delta_{5/2^{+}}(2000).$ This proposal gets further support from
the possible assignment of the other baryon states found in the
approach in the $I=1/2,3/2$ with $J^{P}=1/2^{+},3/2^{+},5/2^+$
sectors to known baryonic resonances. In particular,
$\Delta_{1/2^{+}}(1750)(\ast)$ is naturally interpreted as a $\pi
N_{1/2^{-}}(1650)$ bound state.

\end{abstract}

\maketitle

%%%%%%%%%%%%%%%%%%%%%%%%%%%%%%%%%%%%%%%%%%%%%%%%%%%%%%%%%%%%%%%%%%%%%%

\section{Introduction}

Baryon spectroscopy (see Ref.~\cite{Kle2010} for a recent general
review) is an essential tool to analyze the baryon structure. Data
on baryon masses and transitions, regularly compiled in the Particle
Data Book Review (PDG)~\cite{pdg2010}, allow us when confronted with
theoretical calculations to learn about the effective constituent
degrees of freedom and their interactions inside the baryon. From
the experimental point of view the information on baryonic
resonances mainly comes from pion-nucleon ($\pi N)$ scattering
experiments. The photon nucleon ($\gamma N)$ reactions have led to
advancement in the field, reconfirming many known resonances and
claiming evidence for new ones. From the theoretical point of view
it has become clear in the last years that the primitive quark model
view of a baryon as formed by three effective valence quarks ($3q)$
may require the implementation of higher Fock space terms, in the
form of $4q1\overline{q},$ $5q2\overline{q}$... or meson-baryon,
meson-meson-baryon... components\ to provide a satisfactory
explanation of some baryonic resonances. Paradigmatic cases are the
$\Lambda(1405)$ $S_{01}$ and the $\Delta(1930)$ $D_{35}.$ For
$\Lambda _{1/2^{-}}(1405)$ the relevance of $\overline{K}N$ was
first pointed out in 1977 \cite{Jon77} (more recently the role of
$\pi\Sigma$ has been also emphasized \cite{Mag05}). For
$\Delta_{5/2^{-}}(1930)$ the important role of $\rho\Delta$ has been
recognized \cite{Pej09}. These are particular examples of a more
general situation where a $3q$ model calculation (providing a
reasonable overall description of the whole spectrum) overestimates
the mass of a resonance so that a meson-baryon threshold lies in
between the calculated $3q$ value and the experimental data
\cite{Gon07}. As a consequence, the meson-baryon component may be
dominant when the meson-baryon interaction is attractive, and the
dynamical generation of the resonance from these hadronic degrees of
freedom may be more efficient than a quark model description which
would require $4q1\overline{q}$ and/or higher Fock space terms (note
that the meson and baryon of the threshold might also correspond to
dynamically generated states). This argument can be extended to
resonances where the meson-baryon thresholds are above the $3q$
masses if the meson-baryon interaction is sufficiently attractive as
to provide the binding required by data.

As a matter of fact, meson-baryon components are present in all
baryonic resonances. In some cases the contribution of these
components to the masses may be properly taken into account by
making use of the $3q$ description with effective parameters for the
quark-quark interaction. In other cases, as explained above, this
may not be possible. The intermediate situation corresponds to the
case of resonances for which both approximations may reasonably
reproduce their masses. In such a case the two descriptions may be
at least to some extent equally valid alternatives, the values of
their effective parameters taking implicitly into account the
non-explicit ($3q$ or meson-baryon) component contribution.

In this article we take these considerations into account to analyze
$N$ and $\Delta$ resonances with $J^{P}=1/2^{+},3/2^{+},5/2^{+}$
sectors. The motivation for this study comes mainly from the puzzle
concerning the $\Delta(2000)$ $F_{35}$ ($\ast\ast)$ since the
nominal mass of this resonance does not correspond in fact to any
experimental analysis but to an estimation based on the value of the
masses ($\sim1740$ MeV and $\sim2200$ MeV) extracted from different
data analyses \cite{pdg2010}. This makes feasible the existence of a
hidden $\Delta(1740)$ $F_{35}$ resonance which could not be
reasonably accommodated within a $3q$ framework description what
might be indicating its dynamically generated character. The
theoretical examination of such a possible character is the main
objective of this article. For this purpose we shall follow a
procedure based on the combination of chiral Lagrangians with
nonperturbative unitary techniques in coupled channels of baryons
and/or pseudoscalar and/or vector mesons. This scheme has been very
fruitful in the description of other baryonic resonances through the
analysis of the poles of the meson-baryon or meson-meson-baryon
scattering amplitudes (see for instance Ref.~\cite{meba09} and
references therein).

If existing, the $\Delta(1740)$ $F_{35}$ resonance could be
generated from $\pi-(N(1675)$ $D_{15})$ as suggested in
Ref.~\cite{Gon07}. Since the $N_{5/2^{-}}(1675)$ has been
dynamically generated as a bound state of $\rho\Delta(1232)$ in the
$I=1/2$ sector (the same interaction generating the
$\Delta_{5/2^{-}}(1930)$ for $I=3/2$) \cite{Pej09}, we shall
investigate the three-body $\pi$-$\rho$-$\Delta$ system but keeping
the strong correlations of the $\rho\Delta$ system which generate
the $N_{5/2^-}(1675)$. In such a situation the use of the Fixed
Center Approximation (FCA) to the Faddeev equations is
justified~\cite{Gal:2006cw}. For the sake of consistency, $N$ and
$\Delta$ resonances which can be dynamically generated altogether
with $\Delta_{5/2^{+}}$ will be also analyzed.

The contents of the article are organized as follows. In Section II,
we revisit the cataloged $\Delta_{5/2^{+}}$ resonances and comment
on their $3q$ description. In Section III, we present the FCA
formalism to analyze the $\pi - (\Delta\rho)_{N_{5/2^-}(1675)}$
system. The analysis of the $\pi-(\Delta \rho)_{N_{5/2^-}(1675)}$
scattering amplitude is extracted in Section IV and a tentative
assignment peaks in the amplitudes to baryonic resonances is
proposed. Finally, in Section V we summarize our approach and main
findings.

\section{The $\Delta(2000)$ $F_{35}$ puzzle}

In the PDG \cite{pdg2010} there is only a well established
$\Delta_{5/2^{+}}$ resonance, $\Delta(1905)$ $F_{35}$ ($\ast
\ast\ast\ast),$ and fair evidence of the existence of another one,
$\Delta(2000)$ $F_{35}$ ($\ast\ast)$. However, a careful look at
this last resonance shows that its nominal mass is in fact estimated
from $\Delta (1752\pm32),$ $\Delta(1724\pm61)$ and
$\Delta(2200\pm125)$, respectively, extracted from three independent
analyses \cite{Man92,Vra00,Cut80} of different character: $\pi
N\rightarrow\pi N,\pi\pi N$ in Ref.~\cite{Man92}, multichannel in
Ref.~\cite{Vra00} and $\pi N\rightarrow\pi N$ in Ref. \cite{Cut80}.
Moreover a recent new data analysis has reported a
$\Delta_{5/2^{+}}$ with a pole position at $1738$ MeV \cite{Suz10}.
In this last analysis, incorporating $\pi N,\gamma N\rightarrow\pi
N,\eta N,\pi\pi N$ data, the resonance is obtained from a bare state
at 2162 MeV through its coupling to meson-baryon channels. This bare
state represents the quark core component of the resonance within
this calculation framework.

It is important to remark that i) all the analyses extract the
$\Delta _{5/2^{+}}(1905)$ and ii) the non extraction of
$\Delta_{5/2^{+}}(2200)$ in most of the mentioned analyses may be
related to the restricted range of energy examined (typically below
2200 MeV).

From a $3q$ description the $\Delta_{5/2^{+}}(1905)$ is naturally
accommodated as the lowest $\Delta_{5/2^{+}}$ state in the second
energy band of a double harmonic oscillator potential (one
oscillator for each Jacobi coordinate of the $3q$ system) that
provides (up to perturbative terms) a reasonable overall description
of the whole baryon spectrum \cite{capstick}. Actually quark models
predict two states close in energy for the lowest symmetric and
mixed symmetric orbital configurations in the second energy band.
The $\Delta_{5/2^{+}}(1905)$ is then assigned to the orbitally
symmetric state. Experimental evidence for the mixed symmetric one
has also been reported near $2000$ MeV~\cite{manleyprl1984}.
Similarly, the reported $\Delta_{5/2^{+}}(2200)$, with a more
uncertain mass ($2200\pm125$ MeV), may be reasonably located in the
fourth energy band. On the contrary, $\Delta _{5/2^{+}}(1740)$ lying
far below the energy of the lowest $\Delta_{5/2^{+}}$ state in the
second energy band (the first available band by symmetry to a
$\Delta_{5/2^{+}}$ state) could not be accommodated as a $3q$ state
without seriously spoiling the overall spectral description.

The same kind of problem was tackled in Ref. \cite{Pej09} regarding
the description of $\Delta_{5/2^{-}}(1930)$ with a mass much lower
than the corresponding to the third energy band, the first available
band for such a state. There the consideration of the $\rho\Delta$
channel whose threshold ($2002$ MeV) lies close above the
experimental mass of the resonance and far below the $3q$ mass
($\sim2150$ MeV) allowed for an explanation of
$\Delta_{5/2^{-}}(1930)$ and its partners, $\Delta_{3/2^{-}}(1940)$
and $\Delta_{1/2^{-}}(1900),$ as
$\rho\Delta$ bound states in the $I=3/2$ sector. In addition $N_{1/2^{-}%
}(1650),$ $N_{3/2^{-}}(1700)$ and $N_{5/2^{-}}(1675)$ were also well
described as $\rho\Delta$ bound states in the $I=1/2$ sector,
although the bigger sensitivity in this case to the cutoff parameter
employed left some room for alternative assignments of these
$\rho\Delta$ bound states to nucleonic resonances \cite{Sar10}. It
should be pointed out that, contrary to the $\Delta_{5/2^{-}}(1930)$
and its partners, these nucleon resonances around $1700$ MeV can
also be reasonably described as $3q$ states in the first energy band
\cite{capstick}. Therefore a more reliable explanation of data
should include the contribution of the $3q$ states as well as of the
possible $\rho\Delta$ bound states.

Back to $\Delta_{5/2^{+}}(1740)$ one can easily identify a
meson-baryon threshold, $\left[  \pi N_{5/2^{-}}(1675)\right]
_{\text{threshold}}=1814$ MeV, in between the $3q$ mass ($\sim1910$
MeV) and the data. Then one can wonder about the possibility that
the $\pi N_{5/2^{-}}(1675)$ system may give rise to a bound state
which could provide theoretical support to the fair evidence of the
existence of $\Delta_{5/2^{+}}(1740)$. Actually this bound state
nature could explain why this resonance is extracted in some data
analyses but not in others. It turns out that only analyses
reproducing the $\pi\pi N$ production cross section data extract it.
Let us note that this would be a necessary condition to extract
$\Delta_{5/2^{+}}(1740)$ if corresponding to a $\pi
N_{5/2^{-}}(1675)$ state (let us recall that $N_{5/2^{-}}(1675)$
decays to $\pi N$ and to $\pi\pi N$ with branching fractions of 40\%
and 55\% respectively).

To examine this possibility we perform next an analysis of the $\pi
N_{5/2^{-}}(1675)$ system by assuming that $N_{5/2^{-}}(1675)$ is a
$\rho\Delta$ bound state. Although according to our discussion
above, a combined ($3q+$ $\rho\Delta)$ description of
$N_{5/2^{-}}(1675)$ would be more appropriate we shall consider only
the $\rho\Delta$ bound state option (adequate to our formalism) and
assume that the value of the parameter (cutoff or subtraction
constant) involved in the dynamical generation of
$N_{5/2^{-}}(1675)$ from $\rho\Delta$ takes implicitly into account
the $3q$ component. Furthermore, the same consideration is extended
to the dynamical generation of resonances from $\pi
N_{5/2^{-}}(1675).$

We should finally notice that $\pi N_{5/2^{-}}(1675)$ may couple to
other $s-$wave meson-baryon channel, like
$\pi\Delta_{5/2^{-}}(1930).$ We do not expect this channel to play
any relevant role in the generation of $\Delta_{5/2^{+}}(1740)$
since its threshold is far above in energy. However, the
$\pi\Delta_{5/2^{-}}(1930)$ channel could influence the possible
generation of higher mass resonances. For the sake of simplicity we
shall not include it in our calculation. We should then keep in mind
that the calculated masses for the higher resonances have a higher
degree of uncertainty.

\section{Formalism}

The interaction of a particle with a bound state of a pair of
particles at very low energies or below threshold can be efficiently
and accurately studied by means of the fixed center approximation
(FCA) to the Faddeev equations for the three-particle
system~\cite{chan62}. We shall extend this formalism to include
states above threshold and apply it to
$\pi-(\Delta\rho)_{N_{5/2^-}(1675)}.$ The analysis of the
$\pi-(\Delta\rho)_{N_{5/2^-}(1675)}$ scattering amplitude will allow
us to identify dynamically generated resonances with states listed
in the PDG.

The FCA to the Faddeev equations has been used with success recently
in similar problems of bound three-body systems and contrasted with
full Faddeev or variational calculations. In this sense, the
$N\bar{K}K$ system has been studied with the FCA in
Ref.~\cite{Xie:2010ig}, with very similar results as found in the
full Faddeev calculations in Refs.~\cite{alberto1920} and in the
variational estimate in Ref.~\cite{jido1920}. Similarly, the
study~\cite{Bayar:2011qj} of the $\bar{K}NN$ system within the FCA
has led to very similar results as the variational calculations
of~\cite{Dote:2008hw} when the $\bar{K}N$ chiral amplitudes are
used, or the Faddeev calculation of \cite{Ikeda:2010tk}, when the
energy dependence of the $\bar{K}N$ amplitude is used in agreement
with chiral dynamics.

The important ingredients in the calculation of the total scattering
amplitude for the $\Delta$-$\rho$-$\pi$ system using the FCA are the
two-body $\Delta $-$\rho$, $\Delta$-$\pi$ and $\rho$-$\pi$
unitarized $s-$wave interactions from the chiral unitary approach.
Although the form of these interactions have been detailed elsewhere
\cite{Pej09,Sar05,Roc05}, we shall briefly revisit in Subsection A
the $\Delta$-$\rho$ case. This will allow us to remind the general
procedure of calculating the two-body amplitudes entering the FCA
equations. Then in Subsections B and C the calculation of the $\pi$%
-($\Delta\rho)$ amplitude will be detailed.

\subsection{Unitarized $\Delta\rho$ interaction}

The $\Delta\rho$ interaction has been analyzed in the framework of
the hidden gauge formalism~\cite{gsl} in terms of the exchange of a
$\rho$ meson in the $t-$channel between the $\Delta$ and the $\rho$
\cite{Pej09,Sar10}. Under the low energy approximation of neglecting
$q^{2}/M_{V}^{2}$ in the propagator of the exchanged vector meson,
where $q$ is the momentum transfer, and also the three momentum
$\overrightarrow{k}$ of the vector meson, one obtains for the
$\Delta\rho\rightarrow\Delta\rho$ potential the form
\begin{equation}
V_{pol}=-\frac{1}{4f^{2}}C_{I}(k^{0}+k^{\prime}{}^{0})\vec{\epsilon}\cdot
\vec{\epsilon^{\prime}}, \label{vdeltarho}%
\end{equation}
where $f=93$ MeV is the pion decay constant, $k^{0}(\vec{\epsilon})$ and
$k^{\prime}{}^{0}(\vec{\epsilon^{\prime}})$ is the energy (polarization) of
the incoming/outgoing rho meson and $C_{I}$ an isospin dependent coefficient
with values
\begin{align}
I_{\Delta\rho}=\frac{1}{2}  &  , C_{1/2}=5,\\
I_{\Delta\rho}=\frac{3}{2}  &  , C_{3/2}=2,\\
I_{\Delta\rho}=\frac{5}{2}  &  , C_{5/2}=-3.
\end{align}

Then one can solve the Bethe-Salpeter equation with the on-shell factorized
potential and, thus, the $T$-matrix will be given by
\begin{equation}
T=\frac{V}{1-VG},\label{Tdeltarho}%
\end{equation}
with $V$ the potential of Eq.~(\ref{vdeltarho}) in the isospin basis
without the polarization factor
$\vec{\epsilon}\cdot\vec{\epsilon^{\prime}}$. $G$ is the loop
function for intermediate $\Delta\rho$ states that can be
regularized both with a cutoff prescription as done in
Ref.~\cite{Pej09}, or with dimensional regularization in terms of a
subtraction constant as done in Ref.~\cite{Sar10}. Here we shall
make use of the dimensional regularization scheme better suited to
analyze the sensitivity of our results against variations of the
parameter (small changes of the subtraction constant translate into
significant changes in the values of the cutoff~\cite{Oll01}). The
expression for $G$ is then

\begin{align}
G(s_{\Delta\rho})  &  =i\int\frac{d^{4}q}{(2\pi)^{4}}\frac{2M_{\Delta}%
}{[(P-q)^{2}-M_{\Delta}^{2}+i\epsilon](q^{2}-m_{\rho}^{2}+i\epsilon
)}\nonumber\\
&  =\frac{2M_{\Delta}}{16\pi^{2}}\{a({\mu})+\text{ln}\frac{M_{\Delta}^{2}}%
{\mu^{2}}+\frac{m_{\rho}^{2}-M_{\Delta}^{2}+s_{\Delta\rho}}{2s_{\Delta\rho}%
}\text{ln}\frac{m_{\rho}^{2}}{M_{\Delta}^{2}}\nonumber\\
&  +\frac{\overline{q}}{\sqrt{s_{\Delta\rho}}}[~\text{ln}(s_{\Delta\rho
}-(M_{\Delta}^{2}-m_{\rho}^{2})+2\overline{q}\sqrt{s_{\Delta\rho}%
})+\nonumber\\
&  \text{ln}(s_{\Delta\rho}+(M_{\Delta}^{2}-m_{\rho}^{2})+2\overline{q}%
\sqrt{s_{\Delta\rho}})-\nonumber\\
&  \text{ln}(-s_{\Delta\rho}+(M_{\Delta}^{2}-m_{\rho}^{2})+2\overline{q}%
\sqrt{s_{\Delta\rho}})-\nonumber\\
&  \text{ln}(-s_{\Delta\rho}-(M_{\Delta}^{2}-m_{\rho}^{2})+2\overline{q}%
\sqrt{s_{\Delta\rho}})~]\}, \label{gdeltarho}%
\end{align}
where $P$ is the total incident momentum, which in the center of
mass frame is $(\sqrt{s_{\Delta\rho}},0,0,0)$ being
$\sqrt{s_{\Delta\rho}}$ the invariant mass of the $\Delta\rho$
system. In Eq.~(\ref{gdeltarho}), $\mu$ is the scale of dimensional
regularization and $a({\mu})$ the subtraction constant. Note that
the only parameter dependent part of $G$ is $a({\mu
})+$ln$\frac{M_{\Delta}^{2}}{\mu^{2}}.$ Due to renormalization group
invariance any change in $\mu$ is reabsorbed by a change in
$a({\mu})$ through
$a({\mu}^{\prime})-a({\mu})=$ln$\frac{{\mu}^{\prime2}}{\mu^{2}}$ so
that the amplitude is scale-independent. In Eq.~(\ref{gdeltarho}),
$\overline{q}$ is the momentum of the $\Delta$ or the $\rho$ in the
$\Delta\rho$ center of mass frame, which is given by
\begin{equation}
\overline{q}=\frac{\sqrt{(s_{\Delta\rho}-(M_{\Delta}+m_{\rho})^{2}%
)(s_{\Delta\rho}-(M_{\Delta}-m_{\rho})^{2})}}{2\sqrt{s_{\Delta\rho}}}.
\end{equation}

However, since the $\Delta$ baryon and $\rho$ meson have large total decay
widths $\Gamma_{\Delta}$ and $\Gamma_{\rho}$, they should be taken into
account. For this purpose we replace the $G$ function in Eq.~(\ref{Tdeltarho})
by $\widetilde{G}:$
\begin{align}
\widetilde{G}(s_{\Delta\rho}) &  =\frac{1}{N_{\Delta}N_{\rho}}\int_{M_{\Delta
}-2\Gamma_{\Delta}}^{M_{\Delta}+2\Gamma_{\Delta}}d\widetilde{M}(-\frac{1}{\pi
})\times\nonumber\\
&  \mathcal{I}m\frac{1}{\widetilde{M}-M_{\Delta}+i\frac{\Gamma_{1}%
(\widetilde{M})}{2}}\int_{(m_{\rho}-2\Gamma_{\rho})^{2}}^{(m_{\rho}%
+2\Gamma_{\rho})^{2}}d\widetilde{m}^{2}\nonumber\\
&  \times(-\frac{1}{\pi})\mathcal{I}m\frac{1}{\widetilde{m}^{2}-m_{\rho}%
^{2}+i\widetilde{m}\Gamma_{2}(\widetilde{m})}\nonumber\\
&  \times G(s_{\Delta\rho},\widetilde{M},\widetilde{m}),
\end{align}
with
\begin{align}
N_{\Delta} &  =\int_{M_{\Delta}-2\Gamma_{\Delta}}^{M_{\Delta}+2\Gamma_{\Delta
}}d\widetilde{M}(-\frac{1}{\pi})\mathcal{I}m\frac{1}{\widetilde{M}-M_{\Delta
}+i\frac{\Gamma_{1}(\widetilde{M})}{2}},\nonumber\\
N_{\rho} &  =\int_{(m_{\rho}-2\Gamma_{\rho})^{2}}^{(m_{\rho}+2\Gamma_{\rho
})^{2}}d\widetilde{m}^{2}(-\frac{1}{\pi})\mathcal{I}m\frac{1}{\widetilde
{m}^{2}-m_{\rho}^{2}+i\widetilde{m}\Gamma_{2}(\widetilde{m})},\nonumber
\end{align}
where
\begin{align}
\Gamma_{1}(\widetilde{M}) &  =\Gamma_{\Delta}\left(  \frac{\lambda
^{1/2}(\widetilde{M}^{2},M_{N}^{2},m_{\pi}^{2})2M_{\Delta}}{\lambda
^{1/2}(M_{\Delta}^{2},M_{N}^{2},m_{\pi}^{2})2\widetilde{M}}\right)
^{3}\nonumber\\
&  \times\theta(\widetilde{M}-M_{N}-m_{\pi}),\nonumber\\
\Gamma_{2}(\widetilde{m}) &  =\Gamma_{\rho}\left(  \frac{\widetilde{m}%
^{2}-4m_{\pi}^{2}}{m_{\rho}^{2}-4m_{\pi}^{2}}\right)  ^{3/2}\theta
(\widetilde{m}-2m_{\pi}),\nonumber
\end{align}
with $\lambda(x,y,z)=x^{2}+y^{2}+z^{2}-2xy-2xz-2yz$ the triangle function. We
shall take $\Gamma_{\Delta}=120$ MeV and $\Gamma_{\rho}=150$ MeV.

In addition, one issue worth mentioning is that the spin dependence comes from
the $\vec{\epsilon}\cdot\vec{\epsilon^{\prime}}$ factor of the $\rho$ meson.
The spin of the $\Delta$ baryon does not appear in the present formalism due
to the approximations done. The $\vec{\epsilon}\cdot\vec{\epsilon^{\prime}}$
scalar structure indicate $s-$wave interaction of $\Delta\rho$, therefore, one
has degeneracy for the $J^{P}=1/2^{-},3/2^{-},5/2^{-}$ states for both
$I_{\Delta\rho}=1/2$ and $I_{\Delta\rho}=3/2$.

In order to evaluate the value of the scattering amplitude we have
to fix the parameter $a({\mu})+$ln$\frac{M_{\Delta}^{2}}{\mu^{2}}.$
As explained above the choice of ${\mu}$ is rather arbitrary since a
change in it is reabsorbed by a change in $a({\mu}).$ Values of
${\mu}$ from 630 MeV to 1000 MeV have been employed in the
literature. We choose ${\mu=800}$ MeV, a value rather close to the
cutoff employed in Ref.~\cite{Pej09} ($q_{max}=770$ MeV), and fix
$a(\mu),$ according to our comment at the end of the Section II, to
get the ($\Delta \rho)_{I=1/2}$ bound state at 1675 MeV as
corresponding to the estimated mass of $N_{5/2^{-}}(1675)$ in
Ref.~\cite{pdg2010}$.$ We get $a_{\Delta\rho}=-2.28$ (if instead we
had used ${\mu=630,1000}$ MeV we would have obtained $a_{\Delta
\rho}=-2.76,-1.83).$

\begin{figure}[ptbh]
\includegraphics[scale=0.4]{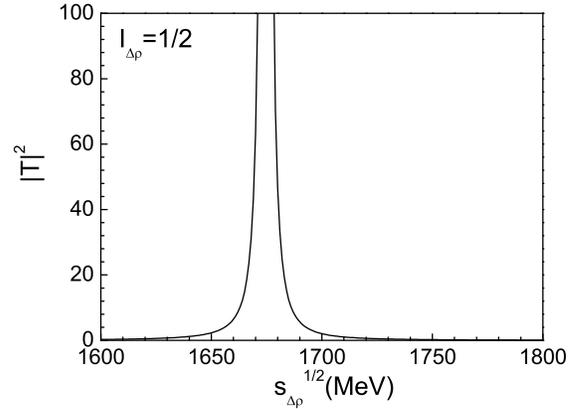}\newline\vspace{-0.5cm}
\caption{Modulus squared of the $\Delta\rho$ scattering amplitude for
$I_{\Delta\rho}=1/2$.}%
\label{Fig:deltarhoTs}%
\end{figure}

In Fig.~\ref{Fig:deltarhoTs} the modulus squared of the scattering
amplitude as a function of the invariant mass of the $\Delta\rho$
system for $I_{\Delta\rho}=1/2$ is shown. Note that in the
$I_{\Delta\rho}=3/2$ sector the bound state is located at
$\sqrt{s_{\Delta\rho}}=1887$ MeV, a little bit lower than its
location in our previous study \cite{Pej09} as a consequence of the
fine tuning of the parameter to get the $I_{\Delta\rho}=1/2$ bound
state at $\sqrt{s_{\Delta\rho}}=1675$ MeV. Notice anyhow that the
assignment of the $I_{\Delta\rho}=3/2$ states at 1887 MeV to
$\Delta(1900)S_{31}(\ast\ast),$ $\Delta(1940)D_{33}(\ast)$ and
$\Delta(1930)D_{35}(\ast\ast)$ remains unambiguous.

\subsection{Single-scattering contribution for the $\pi$ interaction with the
$\Delta\rho$ system}

The FCA to the Faddeev equations for the three body
$\Delta$-$\rho$-$\pi$ system is depicted diagrammatically in
Fig.~\ref{diagram}. The external $\pi$ meson interacts successively
with the $\Delta$ baryon and $\rho$ meson which form the
$N_{5/2^{-}}(1675)$ ($\equiv N^{\ast}$). In terms of two partition
functions $T_{1}$ and $T_{2}$, the FCA equations are
\begin{align}
T_{1}  &  =t_{1}+t_{1}G_{0}T_{2},\label{T1}\\
T_{2}  &  =t_{2}+t_{2}G_{0}T_{1},\label{T2}\\
T  &  =T_{1}+T_{2}, \label{T}%
\end{align}
where $T$ is the total three-body scattering amplitude and $T_{i}$
($i=1,2$)~\footnote{In the present work, the label $1$ represents
$\Delta$ baryon of the compound system, while $2$ represents the
$\rho$ meson.} account for the diagrams starting with the
interaction of the external particle with particle $i$ of the
compound system. Hence, $t_{i}$ represent the $\Delta\pi$ and
$\rho\pi$ unitarized scattering amplitudes whose forms were derived
in Refs.~\cite{Sar05} and \cite{Roc05} respectively to which we
refer for details. In the above equations, $G_{0}$ is the loop
function for the $\pi$ meson propagating inside the
$N_{5/2^{-}}(1675)$ resonance which will be discussed later on.

More specifically $t_{1}$ is the appropriate combination of the $I=1/2,3/2$
and $5/2$ unitarized two-body $\Delta\pi$ scattering amplitudes ($t_{\Delta
\pi}^{1/2}$, $t_{\Delta\pi}^{3/2}$, and $t_{\Delta\pi}^{5/2}$) whereas $t_{2}$
stands for the corresponding combination of the $I=0,1,2$ two-body $\rho\pi$
scattering amplitudes($t_{\rho\pi}^{0}$, $t_{\rho\pi}^{1}$, $t_{\rho\pi}^{2}%
$). For example, let us consider a cluster of $\Delta\rho$ in isospin $I=1/2$,
the constituents of which we call $1$ and $2,$ and the external $\pi$ meson we
call number $3$. The $\Delta\rho$ isospin states are written as
\begin{align}
|\Delta\rho>_{I=1/2,I_{Z}=1/2}  &  =\sqrt{\frac{1}{2}}|(\frac{3}{2}%
,-1)>-\sqrt{\frac{1}{3}}|(\frac{1}{2},0)>+\nonumber\\
&  \sqrt{\frac{1}{6}}|(-\frac{1}{2},1)>,\\
|\Delta\rho>_{I=1/2,I_{Z}=-1/2}  &  =\sqrt{\frac{1}{6}}|(\frac{1}%
{2},-1)>-\sqrt{\frac{1}{3}}|(-\frac{1}{2},0)>+\nonumber\\
&  \sqrt{\frac{1}{2}}|(-\frac{3}{2},1)>,
\end{align}
where the kets on the right hand sides indicate the $I_{z}$ components of the
particles $1$ and $2$, $|(I_{z}^{(1)},I_{z}^{(2)})>$.

\begin{widetext}
\begin{center}
\begin{figure}[ptbh]
\includegraphics[scale=0.8]{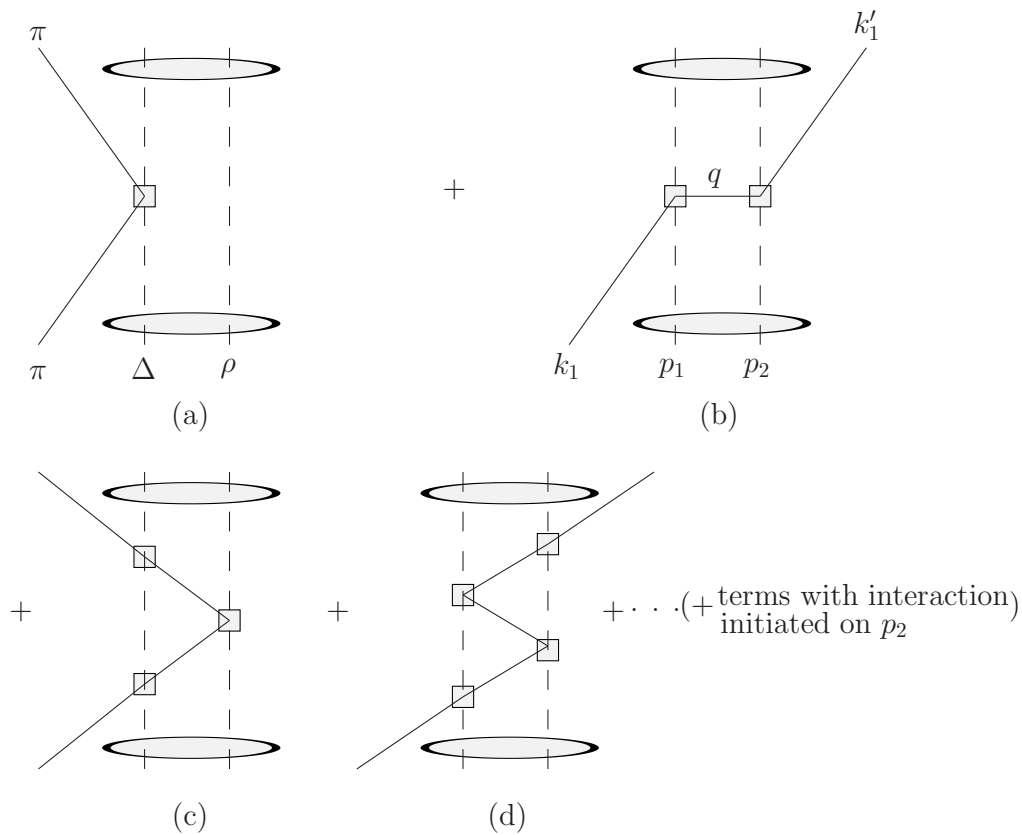}\caption{Diagramatic representation of the
fixed center approximation to the Faddeev equations. Diagrams (a)
and (b) represent the first contributions to the Faddeev equations
from single scattering and double scattering respectively. Diagrams
(c) and (d) represent iterations of the interaction.}
\label{diagram}%
\end{figure}
\end{center}
\end{widetext}

The scattering potential $<\Delta\rho\pi|V|\Delta\rho\pi>$ for the single
scattering contribution (Fig.~\ref{diagram} (a) + term with interaction
initiated on $p_{2})$ can be easily obtained in terms of the two body
potentials $V_{31}$ and $V_{32}$ derived in Refs.~\cite{Sar05} and
\cite{Roc05}.

Here we write explicitly the case of $I_{\Delta\rho}=1/2$ and total
isospin $I_{\Delta \rho\pi}=1/2$,
\begin{widetext}
\begin{eqnarray}
&& <\Delta \rho \pi |V|\Delta \rho \pi > =
(\sqrt{\frac{1}{3}}<(\Delta \rho)|_{I_z=1/2} \bigotimes
<\pi|_{I_z=0} - \sqrt{\frac{2}{3}}<(\Delta \rho)|_{I_z=-1/2}
\bigotimes <\pi|_{I_z=1})
(V_{31}+V_{32}) \nonumber \\
&&(\sqrt{\frac{1}{3}}|(\Delta \rho)>_{I_z=1/2} \bigotimes
|\pi>_{I_z=0} - \sqrt{\frac{2}{3}}|(\Delta \rho)>_{I_z=-1/2}
\bigotimes |\pi>_{I_z=1})
\nonumber \\
&=& ( \sqrt{\frac{1}{3}} (\sqrt{\frac{1}{2}}<(\frac{3}{2}, -1)| -
\sqrt{\frac{1}{3}}<(\frac{1}{2}, 0)| +
\sqrt{\frac{1}{6}}<(-\frac{1}{2}, 1)|) \bigotimes <0| - \nonumber \\
&& \sqrt{\frac{2}{3}} (\sqrt{\frac{1}{6}}<(\frac{1}{2}, -1)| -
\sqrt{\frac{1}{3}}<(-\frac{1}{2}, 0)| +
\sqrt{\frac{1}{2}}<(-\frac{3}{2}, 1)|) \bigotimes <1| )(V_{31}+V_{32}) \nonumber \\
&& ( \sqrt{\frac{1}{3}} (\sqrt{\frac{1}{2}}|(\frac{3}{2}, -1)> -
\sqrt{\frac{1}{3}}|(\frac{1}{2}, 0)> +
\sqrt{\frac{1}{6}}|(-\frac{1}{2}, 1)>) \bigotimes |0> - \nonumber \\
&& \sqrt{\frac{2}{3}} (\sqrt{\frac{1}{6}}|(\frac{1}{2}, -1)> -
\sqrt{\frac{1}{3}}|(-\frac{1}{2}, 0)> +
\sqrt{\frac{1}{2}}|(-\frac{3}{2}, 1)>) \bigotimes |1> ) \nonumber \\
&=& < \frac{\sqrt{10}}{6}((\frac{3}{2} \frac{3}{2}),-1) -
(\frac{\sqrt{15}}{9}(\frac{3}{2} \frac{1}{2}) -
\frac{2\sqrt{3}}{9}(\frac{1}{2} \frac{1}{2})),0) +
(\frac{\sqrt{30}}{18}(\frac{3}{2} -\frac{1}{2})
-\frac{2\sqrt{6}}{9}(\frac{1}{2} -\frac{1}{2})),1)| V_{31} \nonumber \\
&& |  \frac{\sqrt{10}}{6}((\frac{3}{2} \frac{3}{2}),-1) -
(\frac{\sqrt{15}}{9}(\frac{3}{2} \frac{1}{2}) -
\frac{2\sqrt{3}}{9}(\frac{1}{2} \frac{1}{2})),0) +
(\frac{\sqrt{30}}{18}(\frac{3}{2} -\frac{1}{2})
-\frac{2\sqrt{6}}{9}(\frac{1}{2} -\frac{1}{2})),1) > + \nonumber \\
&& < \frac{\sqrt{3}}{6} ((2-1) - (1-1)),\frac{3}{2}) -
(\frac{\sqrt{6}}{6}(20) - \frac{\sqrt{2}}{6} (10)),\frac{1}{2}) +
(\frac{1}{2} (21) - \frac{1}{6} (11)),-\frac{1}{2}) -
\frac{\sqrt{3}}{3} ((22),-\frac{3}{2}) | V_{32} \nonumber \\
&& | \frac{\sqrt{3}}{6} ((2-1) - (1-1)),\frac{3}{2}) -
(\frac{\sqrt{6}}{6}(20) - \frac{\sqrt{2}}{6} (10)),\frac{1}{2}) +
(\frac{1}{2} (21) - \frac{1}{6} (11)),-\frac{1}{2}) -
\frac{\sqrt{3}}{3} ((22),-\frac{3}{2}) >,
\end{eqnarray}
\end{widetext}
where the notation for the states in the third equality is
$((I_{\Delta\pi }I_{\Delta\pi}^{z}),I_{\rho}^{z})$ for the $V_{31}$
matrix element, and $((I_{\rho\pi}I_{\rho\pi}^{z}),I_{\Delta}^{z})$
for the $V_{32}$ one. This leads to the following amplitudes for the
single scattering contribution,
\begin{align}
t_{1}  &  =\frac{5}{9}t_{\Delta\pi}^{I=3/2}+\frac{4}{9}t_{\Delta\pi}%
^{I=1/2}\equiv\frac{5}{9}t_{\Delta\pi}^{3/2}+\frac{4}{9}t_{\Delta\pi}^{1/2},\\
t_{2}  &  =\frac{5}{6}t_{\rho\pi}^{I=2}+\frac{1}{6}t_{\rho\pi}^{I=1}%
\equiv\frac{5}{6}t_{\rho\pi}^{2}+\frac{1}{6}t_{\rho\pi}^{1}.
\end{align}

Proceeding in a similar way, we can get all the amplitudes for the single
scattering contribution required in the present calculation which are shown in
Table~\ref{tab:2bodyamp}.

\begin{table}[ptbh]
\caption{Unitarized two-body scattering amplitudes for the single scattering
contribution. }%
\label{tab:2bodyamp}
\begin{center}%
\begin{tabular}
[c]{|c|c|c|c|}%
$I_{\Delta\rho}$ & $I_{\text{total}}$ & $t_{1}$ & $t_{2}$\\
\vspace*{-0.3cm} &  &  & \\\hline
\vspace*{-0.3cm} &  &  & \\
$\frac{1}{2}$ & $\frac{1}{2}$ & $\frac{4}{9}t_{\Delta\pi}^{1/2} + \frac{5}%
{9}t_{\Delta\pi}^{3/2}$ & $\frac{1}{6}t_{\rho\pi}^{1} + \frac{5}{6}t_{\rho\pi
}^{2}$\\
\vspace*{-0.3cm} &  &  & \\\hline
\vspace*{-0.3cm} &  &  & \\
$\frac{1}{2}$ & $\frac{3}{2}$ & $\frac{1}{36}t_{\Delta\pi}^{1/2} + \frac{2}%
{9}t_{\Delta\pi}^{3/2} + \frac{3}{4}t_{\Delta\pi}^{5/2}$ & $\frac{1}{6}%
t_{\rho\pi}^{0} + \frac{5}{12}t_{\rho\pi}^{1} + \frac{5}{12}t_{\rho\pi}^{2}$\\
&  &  &
\end{tabular}
\end{center}
\end{table}

It is worth noting that the argument of the total scattering amplitude $T$ is
a function of the total invariant mass squared $s$, while the argument in
$t_{1}$ is $s_{1}^{\prime}$ and in $t_{2}$ is $s_{2}^{\prime}$, where
$s_{1}^{\prime}$ and $s_{2}^{\prime}$ are the invariant masses squared of the
external $\pi$ meson with momentum $k_{1}$ and $\Delta$($\rho$) inside the
$N^{*}$ with momentum $p_{1}$($p_{2}$), which are given by
\begin{align}
s_{1}^{\prime}  &  = m^{2}_{\pi} + M^{2}_{\Delta} +\nonumber\\
&  \frac{(M^{2}_{N^{*}}+M^{2}_{\Delta}-m^{2}_{\rho})(s-m^{2}_{\pi}%
-M^{2}_{N^{*}})}{2M^{2}_{N^{*}}},\label{s1prime}\\
s_{2}^{\prime}  &  = m^{2}_{\pi} + m^{2}_{\rho} +\nonumber\\
&  \frac{(M^{2}_{N^{*}}+m^{2}_{\rho}-M^{2}_{\Delta})(s-m^{2}_{\pi}%
-M^{2}_{N^{*}})}{2M^{2}_{N^{*}}}. \label{s2prime}%
\end{align}

Following the approach developed in Ref.~\cite{fcarhorho}, we can easily write
down the $S-$matrix for the single scattering term (Fig.~\ref{diagram} (a) +
term with interaction initiated on $p_{2})$ as,%

\begin{align}
S^{(1)} &  =S_{1}^{(1)}+S_{2}^{(1)}\nonumber\\
&  =((-it_{1})\frac{1}{\mathcal{V}^{2}}\sqrt{\frac{M_{\Delta}}{E_{\Delta}}%
}\sqrt{\frac{M_{\Delta}}{E_{\Delta}^{\prime}}}\frac{1}{\sqrt{2\omega_{\pi}}%
}\frac{1}{\sqrt{2\omega_{\pi}^{\prime}}}+\nonumber\\
&  (-it_{2})\frac{1}{\mathcal{V}^{2}}\frac{1}{\sqrt{2\omega_{\rho}}}\frac
{1}{\sqrt{2\omega_{\rho}^{\prime}}}\frac{1}{\sqrt{2\omega_{\pi}}}\frac
{1}{\sqrt{2\omega_{\pi}^{\prime}}})\nonumber\\
&  F_{N^{\ast}}(\frac{\vec{k_{1}}-\vec{k_{1}^{\prime}}}{2})(2\pi)^{4}%
\delta^{4}(k_{1}+K_{N^{\ast}}-k_{1}^{\prime}-K_{N^{\ast}}^{\prime
}),\label{s12}%
\end{align}
where $\mathcal{V}$ stands for the volume of a box where we normalize to unity
our plane wave states. In Eq.~(\ref{s12}), $F_{N^{\ast}}(\frac{\vec{k_{1}%
}-\vec{k_{1}^{\prime}}}{2})$ is the form factor of the
$N_{5/2^{-}}(1675)$ as a bound state of $\Delta\rho$. This form
factor was taken to be unity neglecting the
$\vec{k},\vec{k^{\prime}}$ momentum in Ref.~\cite{fcarhorho} since
only states below threshold were considered. To consider states
above threshold, we project the form factor into s-wave, the only
one that we consider. Thus

\begin{align}
F_{N^{*}}(\frac{\vec{k_{1}}-\vec{k^{\prime}_{1}}}{2}) \Rightarrow
FFS(s) =\frac {1}{2} \int_{-1}^{1} F_{N^{*}}(k) d(cos\theta),
\end{align}
with
\begin{align}
k=k_{1}\sqrt{\frac{1-cos\theta}{2}},
\end{align}
and
\begin{align}
k_{1}=\frac{\sqrt{(s-(M_{N^{*}}+m_{\pi})^{2})(s-(M_{N^{*}}-m_{\pi})^{2})}%
}{2\sqrt{s}},
\end{align}
is the module of the momentum of $\pi$ meson in the $\pi
N_{5/2^{-}}(1675)$ center of mass frame when $\sqrt{s}$ is above the
threshold of the $\pi N_{5/2^{-}}(1675)$ system, otherwise, $k_{1}$
equals zero. The expression of $F_{N^*}(k)$ is given in the next
section~\footnote{The form factor that we use is suited to a
molecule with two components with equal masses. Some different
recoil corrections are needed when the two masses are
different~\cite{YamagataSekihara:2010yz}, but the results only
affect moderately the peak around $2200$ MeV.}.

\begin{figure}[ptbh]
\begin{center}
\includegraphics[scale=0.4] {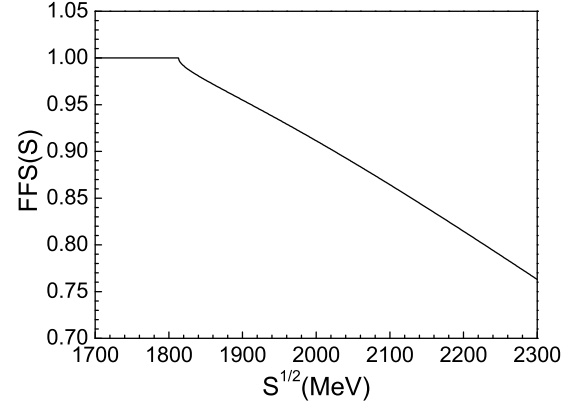}
\end{center}
\caption{Form factor for the single scattering contribution.}%
\label{Fig:ffs}%
\end{figure}

In Fig.~\ref{Fig:ffs}, we show the projection over s-wave of the
form factor for the single scattering contribution as a function of
the total invariant mass of $\Delta\rho\pi$ system.

\subsection{Double-scattering and resummation contribution}

In order to obtain the amplitude of the double-scattering contribution
(Fig.~\ref{diagram} (b) + term with interaction initiated on $p_{2}$) one can
proceed in the same way as in the case of the multi-rho meson interaction in
Ref.~\cite{fcarhorho}. The expression for the $S-$matrix for the double
scattering is ($S_{2}^{(2)}=S_{1}^{(2)}$)%

\begin{align}
S_{1}^{(2)}  &  = (-i t_{1} t_{2}) (2\pi)^{4} \delta^{4}(k_{1}+K_{N^{*}}%
-k_{1}^{\prime}-K_{N^{*}}^{\prime}) \times\nonumber\\
&  \frac{1}{\mathcal{V}^{2}} \sqrt{\frac{M_{\Delta}}{E_{\Delta}}} \sqrt
{\frac{M_{\Delta}}{E_{\Delta}^{\prime}}} \frac{1}{\sqrt{2\omega_{\rho}}}
\frac{1}{\sqrt{2\omega_{\rho}^{\prime}}} \frac{1}{\sqrt{2\omega_{\pi}}}
\frac{1}{\sqrt{2\omega_{\pi}^{\prime}}}\nonumber\\
&  \times\int\frac{d^{3}\vec{q}}{(2\pi)^{3}} F_{N^{*}}(q) \frac{1}{q^{0^{2}%
}-\vec{q}~^{2}-m^{2}_{\pi}+i\epsilon}, \label{s2}%
\end{align}
where $F_{N^{*}}$ is the $N_{5/2^{-}}(1675)$ form factor, and we will take
$q^{0}$ in the $\pi N_{5/2^{-}}(1675)$ center of mass frame, $q^{0} =
(s+m^{2}_{\pi}-M^{2}_{N^{*}})/(2\sqrt{s})$.

Following the approach of Ref.~\cite{danielprd81}, we can get the
expression for the form factor $F_{N^{\ast}}(q)$,
\begin{align}
&  F_{N^{\ast}}(q)=\frac{1}{\mathcal{N}}\int_{|\vec{p}|<\Lambda,|\vec{p}%
-\vec{q}|<\Lambda}d^{3}\vec{p}\frac{M_{\Delta}}{E_{\Delta}(\vec{p})}\frac
{1}{2\omega_{\rho}(\vec{p})}\times\nonumber\\
&  \frac{1}{M_{N^{\ast}}-E_{\Delta}(\vec{p})-\omega_{\rho}(\vec{p}%
)+i(\frac{\Gamma_{\Delta}+\Gamma_{\rho}}{2})}\times\nonumber\\
&  \frac{M_{\Delta}}{E_{\Delta}(\vec{p}-\vec{q})}\frac{1}{2\omega_{\rho}%
(\vec{p}-\vec{q})}\times\nonumber\\
&  \frac{1}{M_{N^{\ast}}-E_{\Delta}(\vec{p}-\vec{q})-\omega_{\rho}(\vec
{p}-\vec{q}) + i(\frac{\Gamma_{\Delta}+\Gamma_{\rho}}{2})}, \label{ff}%
\end{align}
where the normalization factor $\mathcal{N}$ is
\begin{align}
\mathcal{N}  &  =\int_{|\vec{p}|<\Lambda}d^{3}\vec{p}\left(  \frac{M_{\Delta}%
}{E_{\Delta}(\vec{p})}\frac{1}{2\omega_{\rho}(\vec{p})}\right)  ^{2}%
\times\nonumber\\
&  \frac{1}{(M_{N^{\ast}}-E_{\Delta}(\vec{p})-\omega_{\rho}(\vec{p}%
)+i(\frac{\Gamma_{\Delta}+\Gamma_{\rho}}{2}))^{2}},
\end{align}
with $\Gamma_{\Delta}$ and $\Gamma_{\rho}$ the total decay width of
the $\Delta$ baryon and the $\rho$ meson, respectively, taken as in
Subsection A equal to $120$ MeV and $150$ MeV. Since $M_{N^*} <
M_{\Delta}+m_{\rho}$ the effect of the widths of $\Delta$ baryon and
$\rho$ meson is not very important.

To connect with the dimensional regularization procedure we choose
the cutoff $\Lambda$ such that the value of the $G$ function of
Eq.~(\ref{gdeltarho}) at threshold coincides in both methods. Thus
for $\Lambda=820$ MeV we get $M_{N_{5/2^{-}}}=1675$ MeV as required.

\begin{figure}[ptbh]
\begin{center}
\includegraphics[scale=0.4] {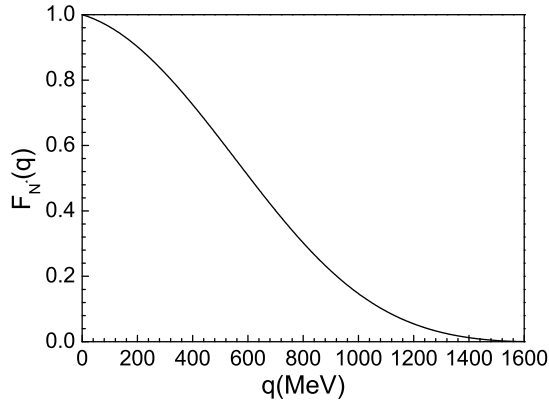}
\end{center}
\caption{Form factor of the $N_{5/2^{-}}(1675)$ as a $\Delta\rho$ bound
state.}%
\label{Fig:ff}%
\end{figure}

We show the form factor $F_{N^{\ast}}(q)$ in Fig.~\ref{Fig:ff} with
$\Lambda=820$ MeV. The condition $|\vec{p}-\vec{q}|<\Lambda$ implies
that the form factor is exactly zero for $q>2\Lambda$. Therefore the
integration in Eq.~(\ref{s2}) has an upper limit of $2\Lambda$.

Before proceeding further, we examine the normalization for the $S$ matrix. We
follow Mandl-Shaw~\cite{mandl} normalization for the fields of baryons and
mesons, then the $S-$matrix for $\pi N^{\ast}$ scattering is written as
\begin{align}
S  &  =-iT_{\pi N^{\ast}}\frac{1}{\mathcal{V}^{2}}\sqrt{\frac{M_{N^{\ast}}%
}{E_{N^{\ast}}}}\sqrt{\frac{M_{N^{\ast}}}{E_{N^{\ast}}^{\prime}}}\frac
{1}{\sqrt{2\omega_{\pi}}}\frac{1}{\sqrt{2\omega_{\pi}^{\prime}}}%
\times\nonumber\\
&  (2\pi)^{4}\delta^{4}(k_{1}+K_{N^{\ast}}-k_{1}^{\prime}-K_{N^{\ast}}%
^{\prime}). \label{s}%
\end{align}

By comparing Eq.~(\ref{s}) with Eq.~(\ref{s12}) for the
single-scattering and Eq.~(\ref{s2}) for the double-scattering, we
see we have to give a weight to $t_{1}$ and $t_{2}$ such that
Eqs.~(\ref{s12}) and (\ref{s2}) get the weight factors that appear
in the general formula of Eq.~(\ref{s}). This is achieved by
replacing
\begin{align}
t_{1}(t_{\Delta\pi})\rightarrow\widetilde{t_{1}}(\widetilde{t_{\Delta\pi}})
&  \equiv t_{1}\sqrt{\frac{M_{\Delta}}{E_{\Delta}}}\sqrt{\frac{M_{\Delta}%
}{E_{\Delta}^{\prime}}}\sqrt{\frac{E_{N^{\ast}}}{M_{N^{\ast}}}}\sqrt
{\frac{E_{N^{\ast}}}{M_{N^{\ast}}^{\prime}}},\nonumber\\
t_{2}(t_{\rho\pi})\rightarrow\widetilde{t_{2}}(\widetilde{t_{\rho\pi}})  &
\equiv t_{2}\frac{1}{\sqrt{2\omega_{\rho}}}\frac{1}{\sqrt{2\omega_{\rho
}^{\prime}}}\sqrt{\frac{E_{N^{\ast}}}{M_{N^{\ast}}}}\sqrt{\frac{E_{N^{\ast}}%
}{M_{N^{\ast}}^{\prime}}}.\nonumber
\end{align}

By solving Eqs.~(\ref{T1}), (\ref{T2}) and summing the two partitions $T_{1}$
and $T_{2}$, we find that
\begin{align}
T_{\pi N^{\ast}} & =\frac{\widetilde{t_{1}}+\widetilde{t_{2}}+2\widetilde
{t_{1}}\widetilde{t_{2}}G_{0}(s)}{1-\widetilde{t_{1}}\widetilde{t_{2}}%
G^{2}_{0}(s)}\nonumber\\
&  +(\widetilde{t_{1}}+\widetilde{t_{2}})(FFS(s)-1),
\end{align}
where $G_{0}(s)$ is given by
\begin{align}
G_{0}(s)  &  =\sqrt{\frac{M_{N^{\ast}}}{E_{N^{\ast}}}}\sqrt{\frac{M_{N^{\ast}%
}}{E_{N^{\ast}}^{\prime}}}\int\frac{d^{3}\vec{q}}{(2\pi)^{3}}F_{N^{\ast}%
}(q)\times\nonumber\\
&  \frac{1}{q^{0^{2}}-\vec{q}~^{2}-m_{\pi}^{2}+i\epsilon}.
\end{align}

\begin{figure}[ptbh]
\begin{center}
\includegraphics[scale=0.4]{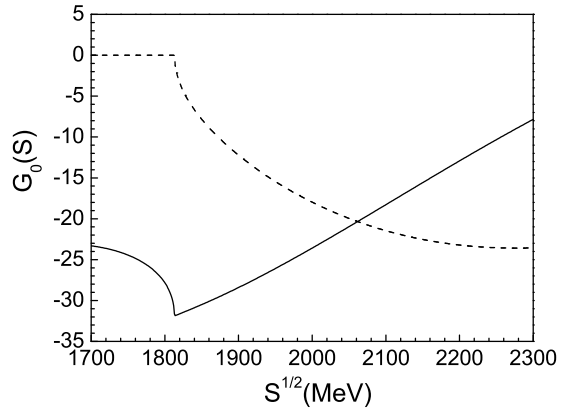}
\end{center}
\caption{Real(solid line) and imaginary(dashed line) parts of the
$G_{0}$
function for the $I_{\Delta\rho}=1/2$.}%
\label{Fig:g01half}%
\end{figure}

In Fig.~\ref{Fig:g01half}, we show the real and imaginary parts of
the $G_{0}$ function for $I_{\Delta\rho}=1/2$ as a function of the
total $\Delta$-$\rho $-$\pi$ system invariant mass.

\section{Results and discussion}

The dynamical generation of resonances from our formalism depends on
three subtraction constants, $a_{\Delta\rho},$ $a_{\Delta\pi}$ and
$a_{\rho\pi},$ respectively associated to the two-body
$\Delta$-$\rho$, $\Delta$-$\pi$ and $\rho$-$\pi$ unitarized $s-$wave
interactions entering our calculation. According to our comments in
Section II we assume they are effective parameters, their values
implicitly taking into account the effects of the $3q$ component of
$N_{5/2^{-}}(1675).$ Following this assumption we have fixed
$a_{\Delta\rho}=-2.28$ in Subsection III.A to get the mass of
$N(1675)$ at its estimated value. Concerning the values of
$a_{\Delta\pi}$ and $a_{\rho\pi}$ they should implement the effect
of the $\pi-(N(1675))_{3q}$ interaction. Therefore they could differ
significantly from the values (around $-2$ for $\mu\sim800$ MeV)
used in the $\Delta\pi$ and $\rho\pi$ calculations of references
\cite{Sar05,Roc05}. In this regard our study has an exploratory
character. We examine first the interval of values of
$a_{\Delta\pi}$ and $a_{\rho\pi}$ around $-2$ allowing for the
dynamic generation of $\pi N(1675)$ bound states in the $I=3/2$
sector. Then we analyze within these intervals the possible
generation of a $\Delta_{5/2^{+}}(1740).$

The general results of this study can be summarized as:

i) the dynamic generation of $I=3/2$ bound states depends essentially on the
value of $a_{\Delta\pi}.$ Only values in the interval $a_{\Delta\pi}<-2.5$
give rise to bound states independently of the value of $a_{\rho\pi}$ (we have
checked this for $-3.0<a_{\rho\pi}<-1.0$ and $-4.5<a_{\Delta\pi}<-2.5).$

ii) if a $I=3/2$ bound state is generated, then two other $I=3/2$
resonances lying above threshold ($1815$ MeV) and below $2300$ MeV
are also generated.

Examples of these results are graphically shown in Fig. 6. In Fig.
6a (6b) the value of $a_{\rho\pi}$ ($a_{\Delta\pi}$) is fixed
whereas $a_{\Delta\pi}$ ($a_{\rho\pi})$ is varied within the
selected interval.

\begin{figure}[ptbh]
\begin{center}
\includegraphics[scale=0.4]{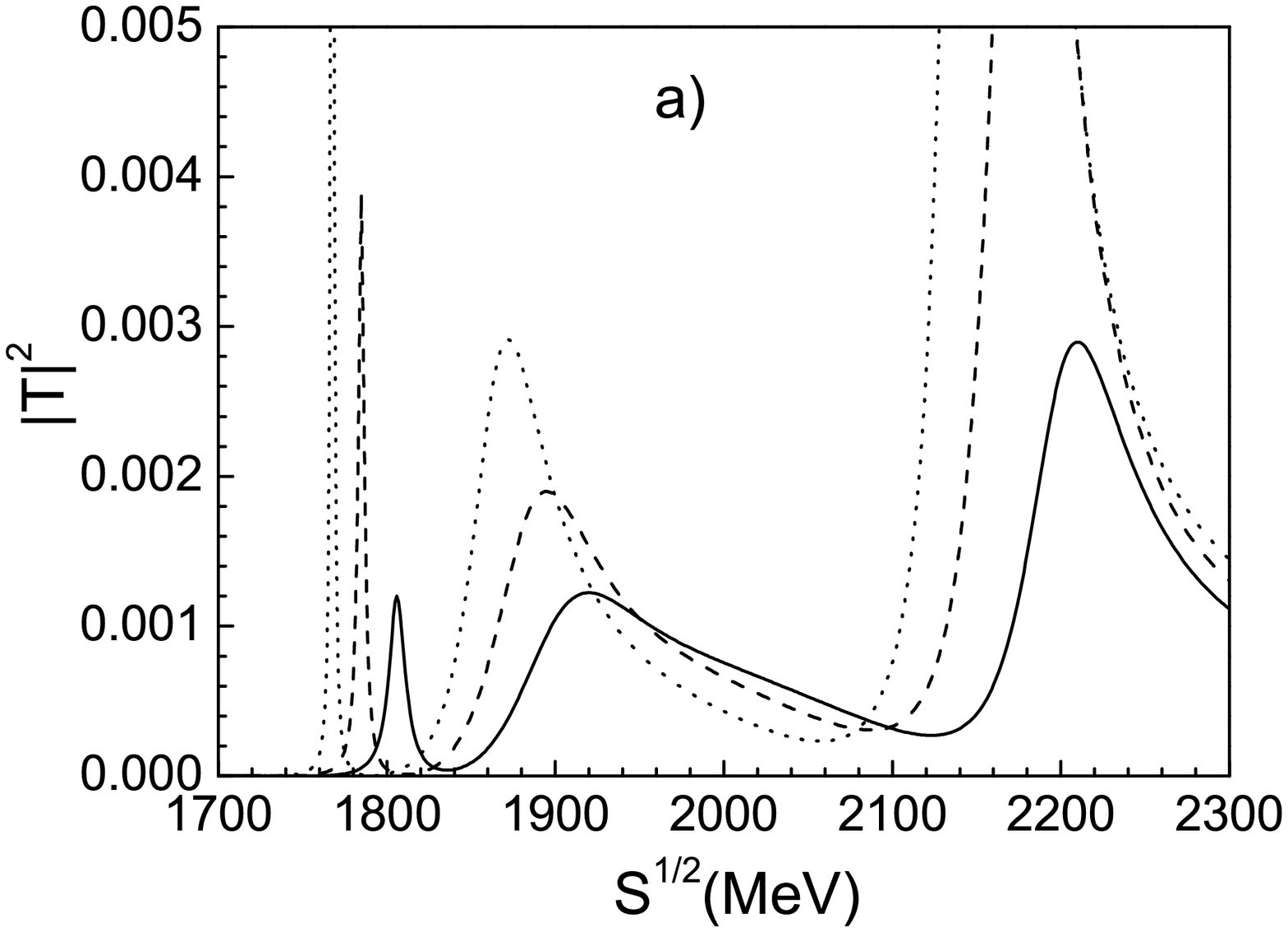} \newline%
\includegraphics[scale=0.4]{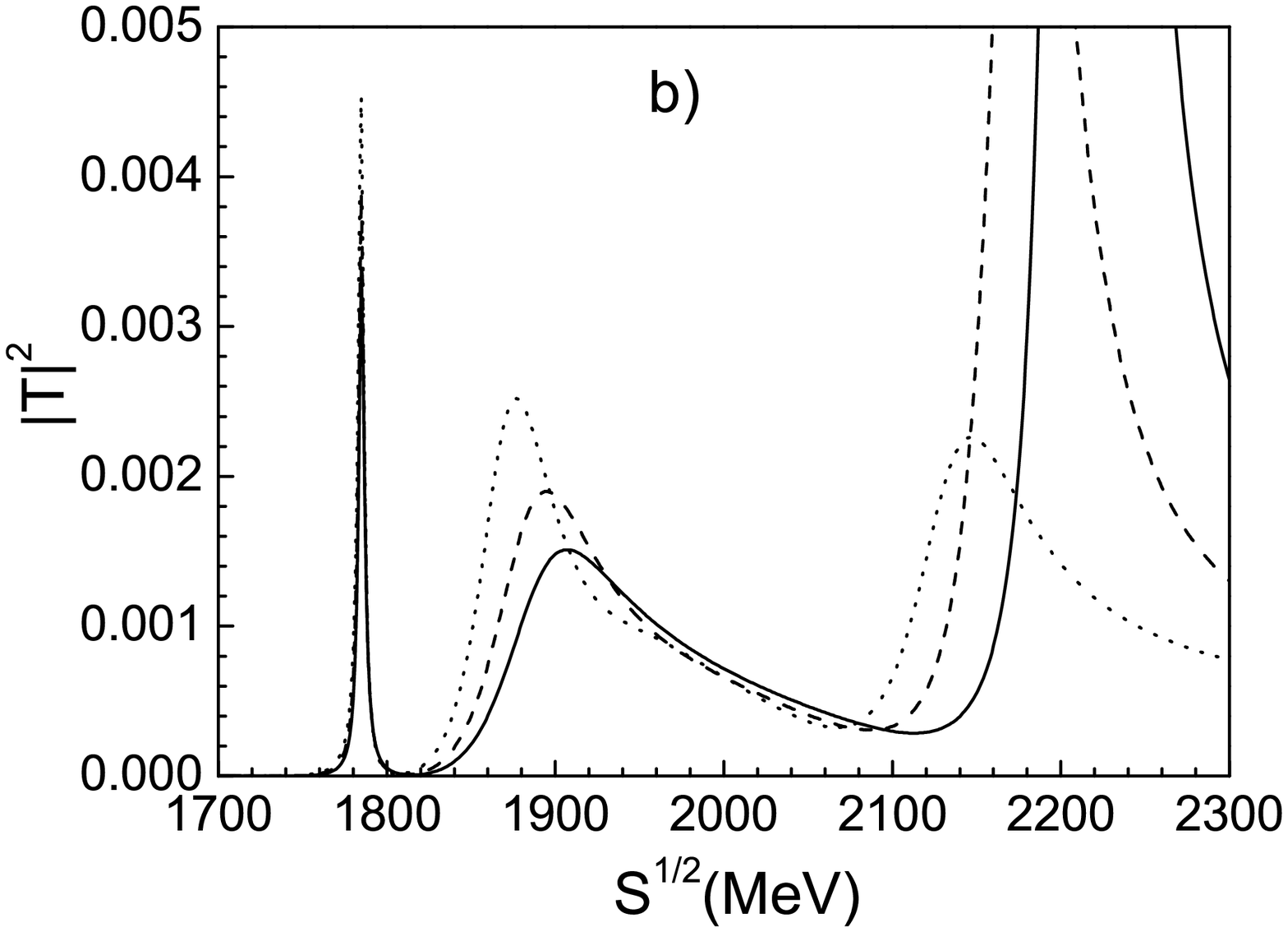}
\end{center}
\caption{Modulus squared of the three-body scattering amplitude for
$I=3/2$. a): results obtained with $a_{\rho\pi}=-2.0$ and
$a_{\Delta\pi }=-2.6$ (solid line), $-3.0$ (dash line), $-3.4$
(dotted line). (b): results obtained with $a_{\Delta\pi}=-3.0$ and
$a_{\rho\pi}=-1.4$ (solid
line), $-2.0$ (dash line), $-2.6$ (dotted line).}%
\label{Fig:3half12}%
\end{figure}

As the positions of the three peaks in the figures are quite stable
(within 60 MeV) against variation of the parameters in the ranges of
values considered, they may be unambiguously assigned to
$\Delta_{5/2^{+}}(1740),$ $\Delta _{5/2^{+}}(1905)$ and
$\Delta_{5/2^{+}}(2200).$ Let us realize that in the region of the
second peak around $2000$ MeV there might be a second resonance, as
reported in Ref.~\cite{manleyprl1984}.

Note that the location of the first peak varies in Fig. 6a from
$1770$ MeV to $1800$ MeV whereas the estimated masses of the
existing candidates in Ref.~\cite{pdg2010}, $\Delta(1752\pm32)$ and
$\Delta(1724\pm61),$ have their upper limits at $1785$ MeV. This
indicates that values $a_{\Delta\pi}\leq-3.0$ can reproduce the
experimental mass. Indeed, we could force $a_{\Delta\pi}=-4.3$ to
get an average mass of $1740$ MeV. Then the second peak appears at
$1830$ MeV and would lie below the estimated mass interval
($1865-1915$ MeV) of $\Delta_{5/2^{+}}(1905).$ However, we should
not forget that the additional consideration of the coupling to
$\pi\Delta_{5/2^{-}}(1930)$ could have more effect on this state as
well as on $\Delta_{5/2^{+}}(2200)$. Concerning the needed values of
$a_{\Delta\pi}$ to get the $\Delta_{5/2^{+}}(1740)$ the important
difference with the reference value $-2$ seems to indicate the
relevant role played by the $3q$ component of $N(1675)$ in the
binding process.

Regarding Fig. 6b, a lack of dependence of the bound state
$\Delta_{5/2^{+}}(1740)$ on $a_{\rho\pi}$ is observed. This means
that all the effect of the $\pi-(N(1675))_{3q}$ interaction in
$\pi-(\Delta\rho)$ is incorporated through $a_{\Delta\pi}.$ The
$\rho$-$\pi$ interaction seems to play a marginal role.

Although encouraging, our results should mainly be interpreted as a
fit to fix the parameters in our formalism. In order to gain
confidence about the possible existence of $\Delta_{5/2^{+}}(1740)$
it becomes essential that further predictions from our formalism
(with no free parameters) are successful in the interpretation of
data. Let us examine the situation with more detail in the
$I=3/2,1/2$ sectors.

\subsection{I=3/2}

$\Delta$ resonances generated from $\pi N_{3/2^{-}}(1700)$ and $\pi
N_{1/2^{-}}(1650)$ are of particular interest since
$N_{3/2^{-}}(1700)$ and $N_{1/2^{-}}(1650)$ are dynamically
generated from $\Delta\rho$ as degenerate states to
$N_{5/2^{-}}(1675).$ As the small mass difference among these
nucleon states ($N^{\ast})$ does not give rise to important mass
differences in the $\pi N^{\ast}$ resonances, we predict
$J^{P}=1/2^{+},3/2^{+}$ experimental $\Delta$ states almost
degenerate to $\Delta_{5/2^{+}}(1740)$, $\Delta_{5/2^{+}}(1905)$ and
$\Delta_{5/2^{+}}(2200).$ Regarding their experimental assignment we
shall centre on possible candidates to be
$\Delta_{3/2^{+},1/2^{+}}(\sim1740)$ and
$\Delta_{3/2^{+},1/2^{+}}(\sim1905)$ since the extensive set of data
available in the energy region below 2.0 GeV makes us confident that
all resonances may have been identified. We shall pay particular
attention to the data analyses of references \cite{Man92} and
\cite{Vra00} extracting both $\Delta_{5/2^{+}}(1740)$ and $\Delta_{5/2^{+}%
}(1905).$ Concerning $\Delta_{3/2^{+},1/2^{+}}(\sim2200)$ they should be
considered as predicted resonances to be extracted when a more complete data
set allows for thorough analyses in the corresponding energy region.

In Table II we list our findings taking for comparison to data the values we
obtain with $a_{\Delta\pi}=-3.4$ and $a_{\rho\pi}=-1.4.$

As can be checked all predicted states can be unambiguously assigned to
experimental resonances.

Particularly interesting is the generation of $\Delta_{1/2^{+}}(1750)$. This
resonance is forced by symmetry to belong to the second energy band and the
quark model overpredicts its mass by about 90 MeV \cite{capstick} (we do not
know any other quark model based on two-quark interactions that does better).
In reference \cite{Gon07} it was argued that it could be generated from $\pi
N_{1/2^{-}}(1650)$ as it is done here (alternatively $\pi\Delta_{1/2^{-}%
}(1620)$ might be also generating it). It should be remarked that only
analyses reproducing the $\pi\pi N$ production cross section data extract it
as it was the case for $\Delta_{5/2^{+}}(1740)$. Therefore the mere existence
of $\Delta_{1/2^{+}}(1750)(\ast)$ could be considered\ within our calculation
framework as an argument in favor of the existence of $\Delta_{5/2^{+}%
}(1740).$

In what respects $\Delta_{3/2^{+}}(\sim1770)$, it is assigned to the
$\Delta_{3/2^{+}}(1600)$, which appears with masses around $1700$
MeV in the analyses of Refs.~\cite{Man92,Vra00}. It should be noted
that its mass is largely overpredicted by $3q$ models as the first
radial excitation of the $\Delta(1232).$ However, other channels as
$\sigma\Delta(1232)$ and $\pi N_{3/2^{-}}(1520)$ could be playing a
more important role in the generation of this resonance.

For the states around $1900$ MeV we should recall that all of them
admit a good $3q$ description. Hence our assignments points out that
an approximately equivalent alternative meson-baryon description is
feasible.

\begin{widetext}
\begin{center}%
\begin{table}[ptbh]
\caption{Assignement of $I=3/2$ predicted states to $J^{P}=1/2^{+}%
,3/2^{+},5/2^{+}$ resonances. Estimated PDG masses for these
resonances as well as their extracted values from references
\cite{Man92} and \cite{Vra00} (in brackets) are shown for
comparison. N. C. stands for a non cataloged resonance in the PDG
review}
\begin{tabular}
[c]{c|ccccc} \hline
Predicted & \multicolumn{5}{c}{PDG data}\\
&  &  &  &  & \\
Mass (MeV) & Name & $J^{P}$ & Estimated Mass (MeV) & Extracted Mass (MeV) & Status\\
1770 & $\Delta(1740)$ & $5/2^{+}$ &  & $1752\pm32$ & N.C.\\
&  &  &  & $(1724\pm61)$ & \\
& $\Delta(1600)$ & $3/2^{+}$ & $1550-1700$ & $1706\pm10$ & *** \\
&  &  &  & $(1687\pm44)$ & \\
& $\Delta(1750)$ & $1/2^{+}$ & $\approx1750$ & $1744\pm36$ & * \\
&  &  &  & $(1721\pm61)$ & \\\hline
$1875$ & $\Delta(1905)$ & $5/2^{+}$ & $1865-1915$ & $1881\pm18$ & ****\\
&  &  &  & $(1873\pm77)$ & \\
& $\Delta(1920)$ & $3/2^{+}$ & $1900-1970$ & $2014\pm16$ & ***\\
&  &  &  & $(1889\pm100)$ & \\
& $\Delta(1910)$ & $1/2^{+}$ &
$1870-1920$ & $1882\pm10$ & ****\\
&  &  &  & $(1995\pm12)$ &
\\\hline
\end{tabular}
\end{table}
\end{center}
\end{widetext}

\subsection{I=1/2}

\begin{figure}[ptbh]
\begin{center}
\includegraphics[scale=0.4]{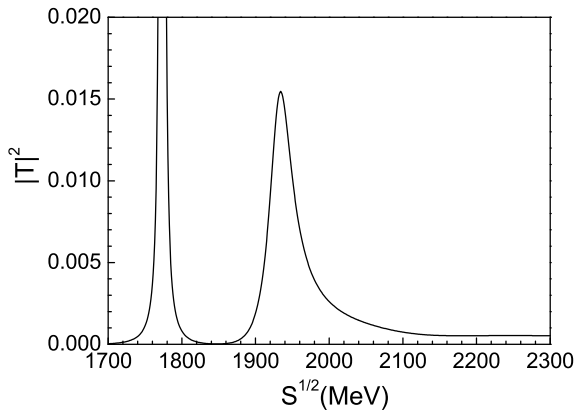}
\end{center}
\caption{Modulus squared of the three-body scattering amplitude for $I=1/2$
with $a_{\Delta\pi}=-3.4$ and $a_{\rho\pi}=-1.4$.}%
\label{Fig:1half3}%
\end{figure}

$N$ resonances are also generated from $\pi N_{5/2^-}(1675)$ and
their partners $\pi N_{3/2^{-}}(1700)$ and $\pi N_{1/2^{-}}(1650).$
In Fig.~\ref{Fig:1half3}, we show the results we get for them with
$a_{\Delta\pi}=-3.4$ and $a_{\rho\pi }=-1.4$ where, as is the
general case in the parameter region explored, there appears two
well defined peaks.

The first peak corresponds to a nucleon resonance almost degenerate
with $\Delta_{5/2^{+}}(1740).$ The mass difference with the second
peak is always about $55$ MeV bigger than that between
$\Delta_{5/2^{+}}(\sim1740)$ and $\Delta_{5/2^{+}}(1905).$

In Table III we show the assignment to experimental states. Again an
unambiguous assignment of predicted states to experimental
candidates can be done. This provides additional support to our
previous predictions. When comparing our results to data we should
be aware, though, that the values used for the parameters have been
fixed from the fitting to $\Delta$ resonances whereas a fitting to
$N$ resonances could give rise to different values of these
parameters. This would be a reflection of the different character of
the $\pi-(N(1675)_{3q}$ and $\pi-(N(1675)_{\rho\Delta}$
interactions. Thus any appreciable deviation of our results from
data could be indicating such a circumstance. This could be indeed
the case in Table III since our predicted masses seem to be
systematically higher than data.

It is worthwhile to recall that $N_{5/2^{+}}(1680),$
$N_{3/2^{+}}(1720)$ and $N_{1/2^{+}}(1710)$ are not well described
by $3q$ models, usually overpredicting their masses by about $70$
MeV. On the other hand other meson-baryon and meson-meson-baryon
channels may be contributing as well to the explanation of these
resonances. For instance $\pi\Delta_{3/2^{-}}(1700)$ may contribute
to $N_{3/2^{+}}(1720)$ and $\pi\Delta_{1/2^{-}}(1620)$ as well as
$\sigma N$ to $N_{1/2^{+}}(1710)$. Indeed in this latter case the
resonance has been dynamically generated as a $\sigma N$ state
\cite{Alb08}.

\begin{widetext}
\begin{center}%
\begin{table}[ptbh]
\caption{Assignement of $I=1/2$ predicted states to $J^{P}=1/2^{+}%
,3/2^{+},5/2^{+}$ resonances. Estimated PDG masses for these
resonances as well as their extracted values from references
\cite{Man92} and \cite{Vra00} (in brakets) are shown for comparison.
N. C. stands for a non cataloged resonance in the PDG review. In
this N. C. case the quoted mass corresponds to reference
\cite{Man92}.}
\begin{tabular}
[c]{c|ccccc}\hline
Predicted & \multicolumn{5}{c}{PDG data}\\
&  &  &  &  & \\
Mass (MeV) & Name & $J^{P}$ & Estimated Mass (MeV) & Mass (MeV) & Status\\
1770 & $N(1680)$ & $5/2^{+}$ & $1680-1690$ & $1684\pm4$ & ****\\
&  &  &  & $(1679\pm3)$ & \\
& $N(1720)$ & $3/2^{+}$ & $1700-1750$ & $1717\pm31$ & ****\\
&  &  &  & $(1716\pm 112)$ & \\
& $N(1710)$ & $1/2^{+}$ & $1680-1740$ & $1717\pm28$ & ***\\
&  &  &  & $(1699\pm 65)$ & \\\hline
$1930$ & $N(2000)$ & $5/2^{+}$ & $\approx2000$ & $1903\pm87$ & **\\
& $N(1900)$ & $3/2^{+}$ & $\approx1900$ & $1879\pm17$ & **\\
& $N(1900)$ & $1/2^{+}$ &  & $1885\pm30$ & N. C.\\\hline
\end{tabular}
\end{table}
\end{center}
\end{widetext}

\section{Summary}

We have performed a Faddeev calculation for the
$\pi-N_{5/2^{-}}(1675)$ system treating the $N_{5/2^{-}}(1675)$ as a
$(\Delta\rho)$ bound state as found in a previous study of the
$\Delta$-$\rho$ system. We have used the fixed center approximation
(FCA) to describe the $\pi-(\Delta\rho)_{N_{5/2^{-}}(1675)}$ system
in terms of the two-body interactions, $\Delta$-$\rho$,
$\Delta$-$\pi$, $\rho$-$\pi,$ provided by the chiral unitary
approach. In order to get a more complete description of
$N_{5/2^{-}}(1675)$ the cutoffs or the subtraction constants $a$
needed to calculate the two-body amplitudes are considered as
effective parameters whose values may implicitly take into account
the effect of the missing $3q$ component of $N_{5/2^{-}}(1675).$
Thus
$a_{\Delta\rho}$ has been fitted to reproduce the nominal mass of $N_{5/2^{-}%
}(1675)$ whereas $a_{\Delta\pi}$ and $a_{\rho\pi}$ are assumed to
incorporate the effects of the $\pi-(N_{5/2^{-}}(1675))_{3q}$
interaction. The variation of the parameters around some values of
reference employed in previous studies of the free $\Delta$-$\pi$
and $\rho$-$\pi$ interactions shows that a quite stable (against
variation of the parameters) bound state is found for
$a_{\Delta\pi}<-2.5$ independently of the value of $a_{\rho\pi}$
what suggests that all the effect of the $3q$ component interaction
can be then absorbed in $a_{\Delta\pi}.$ The significant difference
of the resulting values of $a_{\Delta\pi}$ with respect to the value
of reference seems to indicate the relevance of $3q$ effects.
Indeed, the departure of the subtraction constants $a(\mu)$ from
their natural value is interpreted in Ref.~\cite{Hyodo:2008xr} as a
measure of the relevance of genuine component in the wave function
beyond the meson-baryon ones.

The bound state is always accompanied by the presence of two other resonances
so that a quite precise correspondence to experimental states can be achieved
when the existence of a $\Delta_{5/2^{+}}(1740),$ extracted by two independent
data analyses but non cataloged in the Particle Data Book Review, is taken for
granted. Actually the possibility of providing a theoretical explanation of
such resonance was the main motivation for our study since its description is
clearly out of the scope of the $3q$ model.

Once the parameters are restricted to the bound state region we can generate a
set of definite predictions for $I=3/2,1/2$ and $J^{P}=3/2,1/2.$ All the
generated resonances can be unambiguously assigned to experimental states. It
should be emphasized that this assignment provides a natural explanation to
all the degeneracies observed in the baryonic sectors studied. In particular
it provides theoretical support to the currently poor existence of
$\Delta_{1/2^{+}}(1750)(\ast)$ as an almost degenerate state to $\Delta
_{5/2^{+}}(1740).$ It also points out, confirming previous proposals, the
relevance that meson-baryon components may have in a detailed explanation of
nucleon states as $N_{5/2^{+}}(1680),$ $N_{3/2^{+}}(1720)$ and $N_{1/2^{+}%
}(1710)$ with a deficient $3q$ description.

The consistency of the whole scheme and the good agreement with data
makes us confident that the approximations followed draw the
essential dynamics. From our results we may conclude that there is a
sound theoretical basis to support the data analyses extracting two
distinctive resonances, $\Delta_{5/2^{+}}(1740)$ and
$\Delta_{5/2^{+}}(2200),$ cataloged altogether as $\Delta
_{5/2^{+}}(2000)$ in the Particle Data Book Review. Besides we
predict the existence of $\Delta_{1/2^{+},3/2^{+}}$ resonances about
$2200$ MeV, partners of $\Delta_{5/2^{+}}(2200)$, which may deserve
additional theoretical and experimental analysis.

Concerning $\Delta_{5/2^{+}}(1740)$ (equivalently for
$\Delta_{1/2^{+}}(1750))$ our derivation makes clear its dominant
meson-baryon character so that experimental analyses looking in
detail into specific decay channels would be most welcome. Some of
them ($N\rho$ and $\Delta \pi$~\cite{Vra00}) are already available
in the PDG book. From our theoretical model, a $\Delta \rho$ state
($N_{5/2^-}(1675)$) decays into $\pi N$ and $\pi \Delta$, through a
diagram involving $\rho \to \pi \pi$ with one $\pi$ exchange in the
$t$-channel and the other $\pi$ in the final
state~\cite{Oset:baryon}. This is in agreement with data. Since our
claim for the $\Delta_{5/2^{+}}(1740)$ is a molecular state of $\pi
N_{5/2^-}(1675)$, the natural decay modes would be $\pi \pi N$ and
$\pi \pi \Delta$. Current data in Ref.~\cite{Vra00} suggest that the
$\pi \pi N$ would be the dominant mode.

\section*{Acknowledgments}

This work is partly supported by DGICYT Contract No. FIS2006-03438,
the Generalitat Valenciana in the project PROMETEO, the Spanish
Consolider Ingenio 2010 Program CPAN (CSD2007-00042) and the EU
Integrated Infrastructure Initiative Hadron Physics Project under
contract RII3-CT-2004-506078.

Ju-Jun Xie acknowledges Ministerio de Educaci\'{o}n Grant SAB2009-0116. The
work of A.~M.~T.~is supported by the Grant-in-Aid for the Global COE Program
\textquotedblleft The Next Generation of Physics, Spun from Universality and
Emergence" from the Ministry of Education, Culture, Sports, Science and
Technology (MEXT) of Japan.

P. G. benefits from the funding by the Spanish Ministerio de Ciencia y
Tecnolog\'{\i}a and UE FEDER under Contract No. FPA2007-65748, by the Spanish
Consolider Ingenio 2010 Program CPAN (CSD2007-00042) and by the Prometeo
Program (2009/129) of the Generalitat Valenciana. Partial funding is also
provided by HadronPhisics2, a FP7-Integrating Activities and Infrastructure
Program of the EU under Grant 227431.

\end{document}